\documentclass[pdflatex,sn-mathphys-num]{sn-jnl}% Math and Physical Sciences Numbered Reference Style
\usepackage{graphicx}%
\usepackage{multirow}%
\usepackage{amsmath,amssymb,amsfonts}%
\usepackage{amsthm}%
\usepackage{mathrsfs}%
\usepackage[title]{appendix}%
\usepackage{xcolor}%
\usepackage{textcomp}%
\usepackage{manyfoot}%
\usepackage{booktabs}%
\usepackage{algorithm}%
\usepackage{algorithmicx}%
\usepackage{algpseudocode}%
\usepackage{listings}%
\usepackage{pifont}
\usepackage{colortbl}
\usepackage{array}
\usepackage{float}
\theoremstyle{thmstyleone}%
\theoremstyle{thmstyletwo}%

\theoremstyle{thmstylethree}%

\begin{document}

% "Do LLMs Address the Problems of ABMs? A Critical Review of Generative Social Simulations"

\title[Article Title]{Do Large Language Models Solve the Problems of Agent-Based Modeling? A Critical Review of Generative Social Simulations}

%Do Large Language Models Resolve the Challenges of Agent-Based Modeling? A Critical Review of Generative Social Simulations}

% Validation remains the hard problem for Generative ABMs: A critical review of LLM-based social simulation}
% Do Large Language Models Solve the Problems of Agent-Based Modeling? A Critical Review of Generative Social Simulations} 
%}


\author[1*]{\fnm{Maik} \sur{Larooij}} %\email{m.k.larooij@uva.nl}

\author[1]{\fnm{Petter} \sur{Törnberg}} %\email{p.tornberg@uva.nl}

\affil[1]{\orgname{University of Amsterdam}}
\affil[*]{m.k.larooij@uva.nl}

% \affil*[1]{\orgdiv{ILLC}, \orgname{Organization}, \orgaddress{\street{Street}, \city{City}, \postcode{100190}, \state{State}, \country{Country}}}

% \affil[2]{\orgdiv{Department}, \orgname{Organization}, \orgaddress{\street{Street}, \city{City}, \postcode{10587}, \state{State}, \country{Country}}}

% \affil[3]{\orgdiv{Department}, \orgname{Organization}, \orgaddress{\street{Street}, \city{City}, \postcode{610101}, \state{State}, \country{Country}}}


\abstract{Recent advancements in AI have reinvigorated Agent-Based Models (ABMs), as the integration of Large Language Models (LLMs) has led to the emergence of ``generative ABMs'' as a novel approach to simulating social systems. While ABMs offer means to bridge micro-level interactions with macro-level patterns, they have long faced criticisms from social scientists, pointing to e.g., lack of realism, computational complexity, and challenges of calibrating and validating against empirical data. This paper reviews the generative ABM literature to assess how this new approach adequately addresses these long-standing criticisms. Our findings show that studies show limited awareness of historical debates. Validation remains poorly addressed, with many studies relying solely on subjective assessments of model `believability', and even the most rigorous validation failing to adequately evidence operational validity. We argue that there are reasons to believe that LLMs will exacerbate rather than resolve the long-standing challenges of ABMs. The black-box nature of LLMs moreover limit their usefulness for disentangling complex emergent causal mechanisms. While generative ABMs are still in a stage of early experimentation, these findings question of whether and how the field can transition to the type of rigorous modeling needed to contribute to social scientific theory.}




\maketitle

\section{Introduction}\label{sec1}
Human societies emerge from countless interactions of individuals -- yet, social scientific methods tend to focus on either aggregate patterns or individual decision-making, while neglecting the central question of the relationship between the two \cite{byrne2022complexity}. Agent-Based Models (ABMs) promise a means of changing this, by enabling researchers to simulate how macro-level patterns emerge from micro-level interactions. By modeling individuals as autonomous agents with heterogeneous characteristics and adaptive behaviors, ABMs offer a way to capture the decentralized, dynamic, and non-linear nature of social systems. This approach is particularly appealing for studying phenomena such as collective action, the diffusion of innovations, political polarization, segregation dynamics, and financial market fluctuations -- domains where the emergent dynamics for which traditional analytical methods struggle to account play a central role. ABMs moreover enable exploring how the world could be different, testing how different policy interventions or structural changes might shape societal outcomes.

Despite their theoretical advantages, however, ABMs have historically struggled with widespread adoption within the social sciences. Critics have in particular raised two key objections. 

First, the models have been criticized for oversimplifying human behavior by representing individuals as simple rule-followers or optimizers \cite{xi2023rise,epstein2012generative}. While this approach has been necessary to make computational implementation feasible, it often fails to capture the complexity of human decision-making, which is characterized by more complex reasoning, story-telling, as well as learning, emotions, social norms, and cognitive biases \cite{tornberg2025seeing}. As each model is essentially distinct and a tailor-made representation of a specific social phenomenon \cite{heath2009survey}, comparisons across models tend to be challenging, limiting the possibility for cumulativity of research findings \cite{fagiolo2007critical, windrum2007empirical, ormerod2006validation, epstein1996growing}. 

Second, the models have been criticized for lacking clear connections to the empirical world, often reflecting the assumptions of the modeler more than the realities of human behavior \cite{windrum2007empirical}. Rigorously calibrating and validating the models against real-world data has proven challenging due to the models' complexity and lack of standardization \cite{heath2009survey,windrum2007empirical}. Unlike traditional models with clear parameter estimation techniques, ABMs often involve a large number of assumptions about agent behavior, making empirical validation difficult. Scholars have argued that simulation models must be subjected to rigorous validation if they are to contribute to our understanding of the simulated system \cite{naylor1967verification}. The lack of standardized methods for calibrating these models with real-world data has moreover led to concerns about their reliability and reproducibility \cite{helbing2012social}. Because ABMs can generate highly detailed and complex outputs, distinguishing between meaningful patterns and overfitting to specific assumptions became a major challenge, as the models can be fit to match nearly any data -- in von Neumann's famous phrasing, ``with four parameters I can fit an elephant, with five I can make him wiggle his trunk''. Without rigorous empirical grounding, many ABM studies struggled to move beyond illustrative toy-models, limiting their influence in policy and empirical social science research. 

The models have hence existed in a fundamental tension between the contradictory aims of realism and explainability, and between their inherent versatility and the demands of validation, calibration, and comparability. As a result of these challenges, the data-driven approach of computational social science came to overshadow ABMs from the 2010s onward, following advancement in machine learning and growing availability of digital social data \cite{conte2014agent}.

However, recently, ABMs have made an unexpected comeback. Large Language Models (LLMs) have gained significant traction in recent years, enabling computers to mimic human creativity, reasoning, and language-generation. The rise of LLMs quickly led to the idea that LLMs' capacity to mimic human behavior could be leveraged to simulate social systems, by integrating LLMs with ABMs \cite{tornberg2023simulating,park2023generative}. This promises to solve the challenge of ABMs' lacking realism, by drawing on LLMs' human-like reasoning, language generation, and behavior \cite{guo2024large}. Such ``Generative ABMs'' offer several advantages, allowing agents to be equipped with distinct personalities and human-like capabilities. Generative ABMs have already been harnessed in both cooperative -- achieving a shared goal - and adverserial -- debate and competition -- settings \cite{xi2023rise}, with applications including debating \cite{chan2023chateval}, policy making \cite{xiao2023simulating, hua2023war}, economy \cite{horton2023large, li2023large}, epidemic modeling \cite{williams2023epidemic}, (online) society simulation \cite{park2022social, park2023generative, gao2023s}, psychology \cite{aher2023using, kovavc2023socialai}, gaming \cite{meta2022human, xu2023exploring}, software development \cite{qian2023communicative} and embodied agents \cite{mandi2024roco}.

However, while LLMs promise to address the first key challenge of ABMs by making agents more realistic, their implications for the second -- rigorous validation and calibration -- remain an open question that is central to the future potential of generative ABMs. To address the question of whether generative ABMs enable us to move past the problems that have historically limited the adoption of ABMs, this paper carries out a systematic review of the rapidly growing field of generative ABMs to examine whether and how modelers have dealt with the central challenge of validation. The paper situates generative ABMs in the longer history of ABMs, and the long-standing challenges of validation and calibration. We carry out a systematic review of current research in the field and their validation practices. We find that the issue of validation remains insufficiently addressed in the field, and that it represents a key weakness of the generative ABM literature. Discussing the challenge, we argue that there are reasons to believe that integrating LLMs in ABMs may, in fact, aggravate rather than resolve these challenges, as LLMs are highly complex `black-box' models that intensify the challenges of interpretability, standardization, and replicability \cite{zhao2023explainabilitylargelanguagemodels, luo2024understanding}. 

\section{Agent-Based Modeling in the Social Sciences}
\label{sec:background}
Imagine a murmuration of starlings. Collectively, they form a fluid cloud, moving as a single organism. Yet, there is no ``group mind'' or leader orchestrating the flight. Each bird simply responds to its neighbors, who in turn respond to it -- producing a fluid, nonlinear pattern that appears both choreographed and organic. Modeling this movement from a global perspective -- as, say, a set of interacting variables -- would be impossible. But, as early computer modelers discovered in the 1980s, if we model the individual birds, then the dynamics of the murmuration will simply \textit{emerge} from the bottom-up through the aggregation of local interactions \cite{reynolds1987flocks}.

ABMs take the same approach to studying human social dynamics, treating social systems as consisting of `agents' that perceive their environment, interact and take actions -- and casting society as a form of murmuration: a global pattern that emerges through individual interaction \cite{russell2016artificial}. Traditional sociological methods view society as a hierarchical system where institutions and norms shape individual behavior from the top down, captured through a set of variables. ABMs, by contrast, explore how complex social patterns emerge from decentralized interactions among many agents. Instead of imposing top-down explanations, ABMS view social structures, norms, and institutions as emerging from the interactions of individuals. This shift views human groups as being similar to ant-hills or bird flocks: they are \textit{complex systems}; nonlinear, path-dependent, and self-organizing. and enables studying how these dynamics shape social phenomena -- which is often impossible to achieve using traditional methods \cite{tornberg2025seeing}. ABMs allow examining how macro-patterns may unexpectedly emerge from underlying micro-level mechanisms. At the same time, they only provide \textit{sufficient} explanations, not \textit{necessary} ones, as the same macro-phenomenon can be explained by several different micro-level mechanisms. 

ABM as a social scientific method traces back to the early 1970s, with Thomas Schelling's influential models of social segregation \cite{schelling1971dynamic}. The method however saw rapid growth as the rising availability of computers made modeling accessible to a broad range of researchers. In the 1990s and early 2000s, ABMs grew substantially, together with the broader field of complexity science. The method developed its own journals (e.g., the Journal of Artificial Societies and Social Simulation, JASSS) and  scientific association (The European Social Simulation Association, ESSA).

From 2010 onward, however, ABM and complexity science stalled, challenged by the growth of Computational Social Science and the broader emphasis on data analysis over generative explanation \cite{conte2014agent}.  

ABMs faced growing criticism for being empirically untethered to the systems being represented, reflecting the \textit{ad hoc} intuitions of the modeler rather than the dynamics of the real world. The underlying rules governing the actions of the agents would often be simply assumed by the modeler based on what seemed to them to make sense, often without a clear link to neither empirical data nor existing social scientific theory \cite{windrum2007empirical}. While such an approach may have been acceptable in a situation of data scarcity, the growing availability of digital data in the 2010s intensified demands that ABMs should be calibrated and validated against the real world \cite{naylor1967verification, ormerod2006validation, ziems2024can, macy2002factors}. Critics revived long-standing arguments that models that have not been subjected to proper validation are ``void of meaning'' and contribute nothing to the understanding of the simulated system \cite{naylor1967verification}. Without rigorous empirical grounding, ABMs remain mere toy-models or illustrations, limiting their usefulness for contributing to social science research, and their capacity to influence policy.

As journals increasingly began requiring ABMs to be calibrated and validated against real-world data, this often turned out to be a tall order. As ABMs seek to capture emergent outcomes from large numbers of interacting agents, they are themselves complex systems \cite{silverman2009new}. While this property is necessary for the models to capture social complexity, it also means that the models display many of the pernicious properties that make complex systems challenging to study: the models are often mathematically chaotic, and hence sensitive to initial conditions \cite{bertolotti2020sensitivity}, making them challenging to reproduce. They are moreover often high-dimensional, with their many degrees of freedom resulting in the so-called `curse of dimensionality' \cite{de2005computational}. This dimensionality furthermore makes means that ABMs are computationally costly: the agent interactions tend to scale quadratically with the number of agents, whereas sensitivity analyses scale exponentially with the number of parameters. 

The flexibility and lack of standardization means that few design constraints are imposed by ABMs. As a result, every model is specifically tailored for its intended purpose, making models difficult to compare \cite{fagiolo2007critical}. The lack of comparability represents an impediment to the type of universal methodologies and benchmarks that are crucial for achieving cumulativity of research findings, as it is otherwise unclear how the findings of one model speak to those of another model, let alone to dynamics in the real world. As a result, the field of ABMs faced a ``replication crisis'', as studies found that the models could almost never be replicated -- and that some of the findings in the field turned out to be the results of software bugs \cite{wilensky2007making}.

The lack of empirical grounding implies the need for the models to be highly convincing representations of the systems they simulate. However, social scientists have remained skeptical of the capacity of the often simple rule-based models to represent human behavior. The theoretical descriptions of ABMs as consisting of intelligent, adaptive, and goal-seeking agents that dynamically shape and respond to their environment while taking into account past events \cite{epstein1997artificial} stood in often stark contrast to their realities as simple rule-followers based on a series of ``if-then'' statements or simple optimization. Using such simple models, it is hard to convincingly argue that the models encompass the full complexity of human decision-making, in particular as the role of beliefs, memories, story-telling, and past interactions are rarely accounted for \cite{sargent2010verification}.  While ABMs do represent improvements over the atomized perfect optimizers of neoclassical theory, they remain part of the same broader methodologically individualist paradigm \cite{tornberg2018complex,byrne2022complexity}.

Many social scientists have furthermore found it challenging to integrate ABMs with existing social scientific theory. In part due to their simplicity, the explanations offered by ABMs often seemed to be fundamentally at odds with existing social scientific explanations, as they cast any social phenomenon as simply `emerging' from bottom-up interactions \cite{macy2002factors}. Such reductive explanations tend to come with a hint of the same political biases as neoclassical models, through the lens of which the market appears as `natural', while the top-down role of institutions are viewed as imposing unnatural constraints \cite{baker2022ultimate}. The widely lauded Schelling \cite{schelling1971dynamic} segregation model, for instance, seemed to suggest that segregation naturally ``emerged'' from an innate mutual dislike of the opposing group, while disregarding dominant explanations like structural racism, white flight, and red-lining -- thereby also seeming to eradicate the possibility of collective solutions to address the problem.

As a result of these challenges, together with the growing capacities of machine learning and AI to produce insights from the increasingly abundant digital social data, the social data analytics and computational social science came to overshadow ABMs from the 2010s onward \cite{conte2014agent}. 

\subsection{Generative ABMs}
The emergence of Large Language Models (LLMs) has recently marked an unexpected shift in the fortunes of ABMs. While ABM research had already experimented with using deep learning-based approaches to modeling human behavior \cite{xi2023rise}, LLMs offer an entirely new approach to represent human behavior. As LLMs are trained to mimic human language and reasoning, they appear in many ways to be out-of-the-box models of humans. In contrast to traditional ABMs \cite{zhao2023depth}, such ``generative agents'' are trained on vast amounts of information, giving them an internal world-model that allows them to generalize to new problems and mimic human reasoning. Generative agents can furthermore understand and generate natural language in ways indistinguishable from humans \cite{jones2024people}.

These capacities were first explored in a 2023 paper using LLMs to simulate human behavior in an The Sims-like artificial world, finding that the generative agents would begin to reproduce human-like behavior and social dynamics \cite{park2022social}. Following this, the field of generative ABMs exploded in growth, making it subject to several reviews seeking to survey the emerging field \cite{guo2024large, xi2023rise, wang2024survey, cheng2024exploring, mou2024individual}. 

The agents used in the ABMs are often provided with distinct personas, consisting of, for example, demographic, personal and social information, that is either hand-crafted, generated by AI, or based on real-world datasets. The agents are often given memory modules in which meaningful information can be stored and retrieved. Planning modules allow agents to reason and act according to self-generated plans. Based on these modules, agents are able to take actions to change the state of the environment (e.g. changing the state of a toilet to `occupied'), alter their internal states, or communicate with other agents. 

Generative agents thus seem to promise to resolve the problem of ABMs' lacking realism, by making it trivial to create simulations with agents that can remember, reason, argue, engage in realistic conversations, and achieve ``believable human behavior'' \cite{park2022social}. The substantial excitement surrounding generative ABMs is hence understandable. 

At the same time, it would seem that the question of calibration and validation that has long haunted ABMs remains potentially exacerbated by the introduction of LLMs. We can point to three interrelated ways in which LLMs may aggravate the existing challenges of ABM validation.

First, LLMs are `black-box' models: their capacities are emergent, and it is virtually impossible to understand why a particular model gives a particular output. Language models rely on large-scale neural networks, and their output is the result of billions of parameters \cite{zini2022explainability}.  The more advanced the network gets, the less we know exactly what is going on in its decision-making process. LLMs are furthermore stochastic -- the same input can lead to different outputs -- making validating and reproducing results potentially challenging.

Second, the models cannot be trusted to accurately represent social groups or impersonate human behavior. While models are given descriptions of individuals to impersonate, studies have found that the models are poor at representing groups and their attributes, often engaging in exaggerated stereotypes rather than accurate representations. This is intimately associated with the hotly debated issue of LLM bias \cite{ferrara2023should}.  Specifically, we can point to two types of bias that pose challenges to validation of generative ABMs: \textit{social bias} and\textit{ selection bias} \cite{navigli2023biases}. Social bias refers to the discrimination, stereotypes, or prejudices against certain groups of people, which may mean that models do not accurately represent external groups, but rather engage in problematic stereotypes. Selection bias refers to the choice of texts that make up a training corpus, which means that LLMs may reproduce historical events or complete social network interactions, basing simulations on prior knowledge instead of dynamically producing the dynamics as intended by the modelers. 

Third, as LLMs are probabilistic next-word predictors, they possess no internal mechanism to validate the correctness of their outputs. They may hence generate outputs that are factually incorrect or nonsensical -- a phenomenon referred to as `hallucination' \cite{huang2023survey, zhang2023siren}. This could potentially lead to unexpected effects, or hard-to-solve problems in the internal workings of models, as the outputs may not be coherent with agents' memories or personas (e.g. an agent `recalling' events that it never experienced), leading to unexpected outcomes. 

\subsection{Validation in ABMs}
Before turning to reviewing how validation has been handled in the field of Generative ABMs, we will first draw on the literature on validation in ABMs to provide a classification of types of validation that have been discussed and applied. Validation is the process of assessing the degree to which a model is an accurate representation of the real world \cite{ormerod2006validation}.  Sargent \cite{sargent2010verification} argues that the aim of validation should be `operational validity': ``determining whether the simulation model's output behavior has the accuracy required for the model's intended purpose over the domain of the model's intended applicability''. Models are always abstractions, and it is necessary to bracket part of the system being modeled. `Realism' is, in other words, neither achievable nor desirable. 

First, the literature distinguishes between \textit{internal} and \textit{external} validation methods. External validation makes use of external resources for validation, such as real-world data, well-studied behavior, or humans for judgment. Internal validation is carried out to validate the internal consistency or to do internal observations of the model, for example by monitoring the system and reasoning about the observations or performing a sensitivity analysis \cite{christopher2002identification}.

Second, we can distinguish between `subjective' and `objective' validation. The former includes techniques like human judgment or basic observations of the system in the form of traces, graphics and performance indicators. One influential example of subjective validation is the Turing Test, in which individuals are asked if they can discriminate between human and model outputs \cite{turing2009computing}. Objective validation encompasses techniques such as comparing with real-world data or similar models with the use of statistical tests. One of the earliest papers on model validation proposes a multi-stage validation approach of first developing the model's assumptions, then validating where possible by empirically testing them, and lastly comparing the input-output relations of the model to the real system \cite{naylor1967verification}.

These two dimensions allow us to divide validation techniques into four categories. For example, using human judges would be \textit{subjective} and \textit{external}, whereas a sensitivity analysis would be \textit{objective} and\textit{ internal}.

\section{Method}
\label{sec:design}
% Section dedicated to document your study was carried out

%\subsection{Research Goal}
%The goal of this paper is to provide a clear and organized view of validation practices in the field of generative agent research for world simulations. Furthermore, it identifies challenges that arise in model validation, particularly related to LLMs, and assesses the awareness of these challenges in the field.

To identify papers, we used as a starting-point a series of previous survey studies \cite{guo2024large, xi2023rise, wang2024survey, cheng2024exploring, mou2024individual}, and employed backward snowballing as described in \cite{wohlin2014guidelines}. We furthermore carried out a search to identify relevant new papers using Scopus, a widely used bibliographic database. The search aimed to capture key contributions across disciplines, including computer science, artificial intelligence, and computational social science. We developed a Boolean search query using keywords and phrases that reflect the core topics of interest. The query was designed to be broad enough to capture diverse research while ensuring specificity by targeting the title, abstract, and keywords fields: {\begin{verbatim} TITLE-ABS-KEY ( ( "generative social simulation" ) OR ( "generative 
agent-based model*" ) OR ( "agent-based simulation" AND "generative 
AI" ) OR ( "LLM*" AND "agent-based model*" ) OR ( "large language 
model*" AND "ABM" ) OR ( "foundation model*" AND "ABM" ) OR ( "multi-
agent system*" AND "generative AI" ) OR ( "generative agent*" ) 
OR ( "social simulation" AND "LLM*" ) OR ( "large language model-based 
agents" ) ) \end{verbatim}}

The final search was conducted on 2025-03-27, yielding 35 results. To ensure relevance, we followed a two-step screening process:

\begin{enumerate}
    \item  Title and abstract screening: We removed irrelevant papers based on a preliminary review of titles and abstracts.
    \item Full-text review: The remaining papers were assessed for methodological rigor and direct relevance to the survey's research questions.
\end{enumerate}

We focus on papers focused on using generative ABMs for social simulation. Studies were included if they met the following eligibility criteria: 
\begin{itemize}
\item Papers that use Large Language Models (LLMs) as the basis of a simulated agent
\item Papers that leverage multiple agents
\item Papers that in which agents interact, meaning that their actions depend on the state of the system or on the actions of other agents.
\item Papers in which the model is seen as representing human behavior
\end{itemize}

\noindent Studies were excluded if they:
\begin{itemize}
    \item Focus on task completion rather than on modeling a social phenomenon.
    \item Do not seek to model or achieve similarity with human behavior.
    \item Were not published in English.
    \item Surveyed existing studies rather than presenting new research. 
\end{itemize}

%to model multiple interacting agents with the aim of simulating a social system. We excluded papers that focused on, for instance, individual agent simulation or using LLM agents for task completion, or that do not aim to model or achieve similarity with human behavior. 

\noindent These criteria meant that we excluded papers in which agents made individual decisions, but with these never being affected by other agents, such as Feng et al's \cite{feng2024large} study of human characteristics on electric vehicle charging behavior, or Ren et al's \cite{ren2024bases} study using agents with different personalities to replicate human web search behavior. We also excluded studies using agents to produce synthetic survey data \cite{argyle2023out}, %During the search, some papers were found that took more of an exploration stance towards the system. For example \cite{chuang2023simulating} focused more on whether LLMs can be used to support sophisticated simulation of agent interactions, and in GovSim \cite{piatti2024cooperate}, the goal was to explore the strategic interactions and cooperative decision-making in generative agents. 
and papers aimed at creating frameworks for generative ABMs rather than replicating human behavior, like CAMEL \cite{li2023camel}, MetaGPT \cite{hong2023metagpt}, CGMI \cite{jinxin2023cgmi} and AutoGen \cite{wu2023autogen}. 

% The following selection criteria were applied in the curation of research papers. 
% \begin{enumerate}
% \item[{I}1-] Papers that use Large Language Models (LLMs) as the basis of a simulated agent
% \item[{I}2-] Papers that leverage multiple agents
% \item[{I}3-] Papers that leverage agents that take actions based on the shared state of the system or on the actions of other agents
% \item[{I}4-] Papers that make the explicit or implicit claim that the model is representing human behavior
% \item[{E}1-] We do not include papers that focus on efficient task completion rather than on modeling the a social phenomenon.
% \item[{E}2-] Papers that do not have the validation of the alignment with human behavior as goal
% \end{enumerate}

To extract the needed data, the papers were read in-depth, in particular sections containing the relevant information, such as the introduction, evaluation, and limitations sections. The identified records were inserted into a structured database containing all the classifications.

We are seeking to provide an overview of what systems researchers are seeking to reproduce using generative ABMs, and how they approach validation. We will then draw on this to discuss the current role of the issue of validation within generative ABM research. Our review hence focuses on the following questions: 

\begin{enumerate} 
    \item RQ1: What social phenomena are generative ABMs seeking to reproduce?
    \item RQ2: What strategies are employed to validate the models against the studied phenomena?    
\end{enumerate}

\subsection{The target system}
To answer RQ1, we distinguish between simulating\textit{ individual} and \textit{group} behavior. Papers that seek to reproduce individual behavior focus on the actions of a single agent, for example if decisions are congruent with a given profile. Other papers search for group behavior at the macro-level of the simulation, like the formation of the network through connections, or the propagation of information through the network. For individual behavior, we further distinguish between the following aspects of the simulated phenomenon, which were identified inductively through an initial analysis:

\begin{enumerate}
    \item Alignment with profile: the alignment of the agent's actions with their personality, memories and profile, as well as the specific situation.
    \item Emotion: the emotions expressed by the agent given a specific situation or action.
    \item Conversations and content generation: the communication and dialogue between agents.
    \item Social awareness: the awareness of others and of social norms.
    \item Decision making and reasoning: the decision making and reasoning capabilities.
    \item Opinion and attitude: the opinion or attitude towards some event, and possibly the shift in opinion or attitude.
\end{enumerate}

\noindent For group behavior, we used the following three categories:

\begin{enumerate}
    \item Network propagation: the propagation of something through the network, such as (mis)information, emotion, attitudes, or cultural elements. 
    \item Network structure: the structure of the network, for example the network degree distribution, density or some other network structure attribute.
    \item Social dynamics: phenomena that emerge from individuals actions, for example a 'herd effect'
\end{enumerate}

\noindent Some papers focus on multiple phenomena are thus given multiple categorizations. A paper is classified as focusing on a certain type of behavior only if the paper sets out to model that behavior, thus excluding observed phenomena that are noted only as an aside. For example, Kaiya et al \cite{kaiya2023lyfe} set out to measure social reasoning, but note some human-like emotions and reactions.

\subsection{Validation Techniques}
For RQ2, we draw on existing categorizations in the literature on ABM validation. We hence separate between \textit{internal} and \textit{external} validation, as well as between \textit{subjective} and\textit{ objective} validation. We expand on the resulting four categories based on the initial analysis, and arrive at the following five categories of validation:

\begin{enumerate}
    \item Validation based on human(-like) judgment [External / Subjective]\textbf{:} using input from researchers, experts, crowd-workers or even LLMs to judge the alignment of the model with human behavior
    \item Validation against well-known social patterns\textbf{ }[External / Subjective]\textbf{:} comparing observations of the model to well-established phenomena in human behavior, although not quantitatively
    \item Validation against similar models [External / Subjective or Objective]\textbf{:} comparing subjectively or objectively with the results and behavior of earlier models, such as traditional ABMs.
    \item Validation against human-generated data [External / Objective]\textbf{:} comparing to collected human behavioral data, such as social media interactions, human annotations or earlier experiments  
    \item Validation based on internal consistency [Internal / Objective]\textbf{:} evaluating the internal coherence and consistency of the system, for example through sensitivity analysis.
\end{enumerate}

\noindent We distinguish between primary and secondary validation techniques. Primary techniques are the most important in the validation process of the paper, while secondary techniques are used merely to lend supportive evidence.

\section{Results}
\label{sec:results}

We identify 35 papers that fulfill the eligibility criteria. The full classification of all 35 papers is presented in table \ref{tab:rq1}. The generated content and conversations, social dynamics and network propagation are the most popular areas of study. 21 papers focus on a single category, the others measure multiple categories in one paper. On average, a paper studies 1.63 distinct categories. 

\subsection{Target systems of simulations}
We will begin by reviewing the literature by focusing on what social phenomena that generative ABMs are used to reproduce. We will here discuss each category in turn, thereby providing an overview of the studied phenomena and their validation.

\begingroup
    \renewcommand*{\arraystretch}{1.5}%
    \definecolor{tabred}{RGB}{230,36,0}%
    \definecolor{tabgreen}{RGB}{0,116,21}%
    \definecolor{taborange}{RGB}{255,124,0}%
    \definecolor{tabbrown}{RGB}{171,70,0}%
    \definecolor{tabyellow}{RGB}{255,253,169}%
    \newcommand*{\redtriangle}{\textcolor{tabred}{\ding{115}}}%
    \newcommand*{\greenbullet}{\textcolor{tabgreen}{\ding{108}}}%
    \newcommand*{\orangecirc}{\textcolor{taborange}{\ding{109}}}%
    \newcommand*{\headformat}[1]{{\small#1}}%
    \newcommand*{\vcorr}{%
      \vadjust{\vspace{-\dp\csname @arstrutbox\endcsname}}%
      \global\let\vcorr\relax
    }%
    \newcommand*{\HeadAux}[1]{%
      \multicolumn{1}{@{}r@{}}{%
        \vcorr
        \sbox0{\headformat{\strut #1}}%
        \sbox2{\headformat{Complex Data Movement}}%
        \sbox4{\kern\tabcolsep\redtriangle\kern\tabcolsep}%
        \sbox6{\rotatebox{45}{\rule{0pt}{\dimexpr\ht0+\dp0\relax}}}%
        \sbox0{\raisebox{.5\dimexpr\dp0-\ht0\relax}[0pt][0pt]{\unhcopy0}}%
        \kern.75\wd4 %
        \rlap{%
          \raisebox{.25\wd4}{\rotatebox{45}{\unhcopy0}}%
        }%
        \kern.25\wd4 %
        \ifx\HeadLine Y%
          \showthe\dimen0 % Debugging: Check the computed width
          \dimen0=\dimexpr\wd2+.5\wd4\relax
          \rlap{\rotatebox{45}{\hbox{\vrule \vrule width 3cm height .4pt}}}%
        \fi
      }%
    }
    \newcommand*{\head}[1]{%
  \HeadAux{\def\HeadLine{Y}#1}%
}
    \newcommand*{\headNoLine}[1]{\HeadAux{\global\let\HeadLine=N#1}}%
    \noindent
    \begin{table*}[]
    \footnotesize
    \begin{tabular}{%
      >{\bfseries}lc|
      *{5}{c|}c>{\quad}c|
      *{2}{c|}c>{\quad}c
    }%
      &
      \head{Profile Alignment} &
      \head{Emotion} &
      \head{Conversation / Content} &
      \head{Social Awareness} &
      \head{Decisions / Reasoning} &
      \head{Opinion / Attitude}
      &
      &
      \head{Network Propagation} &
      \head{Network Structure} &
      \head{Social Dynamics} 
      &
      \\
      \sbox0{S}%
      \rule{0pt}{\dimexpr\ht0 + 2ex\relax}%
      Generative Agents \cite{park2023generative} & \greenbullet &&
        & & & &&
      \greenbullet & \greenbullet & 
      \\\hline
      WarAgent \cite{hua2023war} & \greenbullet &&
      & & & &&
       &  \greenbullet &
      \\\hline
      Social Simulacra \cite{park2022social} & && \greenbullet
        & & & &&
       &  & 
      \\\hline
      S3 \cite{gao2023s} & & \greenbullet & \greenbullet & & \greenbullet & \greenbullet & & \greenbullet & &
      \\\hline
      De Marzo et al. \cite{de2023emergence} & & & & & & & & & \greenbullet &
      \\\hline
      Humanoid Agents \cite{wang2023humanoid} & & \greenbullet & \greenbullet & & & \greenbullet & & & &
      \\\hline
      Lyfe Agents \cite{kaiya2023lyfe} & & & & & \greenbullet & & & & &
      \\\hline
      Zhang et al. \cite{zhang2023exploring} & & & & & & & & & & \greenbullet
      \\\hline
      AgentVerse \cite{chen2023agentverse} & & & & & & & & & & \greenbullet
      \\\hline
      Williams et al. \cite{williams2023epidemic} & & & & & \greenbullet & & & & &
      \\\hline
      Project Sid \cite{al2024project} & \greenbullet & & & \greenbullet & \greenbullet & & & \greenbullet & &
      \\\hline
      Li et al. \cite{li2023theory} & & & & \greenbullet & & & & & &
      \\\hline 
      Li et al. \cite{li2024large} & & & & & & & & \greenbullet & &
      \\\hline
      SpeechAgents \cite{zhang2024speechagents} & & & \greenbullet & & & & & & &
      \\\hline
      FPS \cite{liu2024skepticism} & & & & & & & & \greenbullet & & 
      \\\hline
      SOTOPIA \cite{zhou2023sotopia} & & & & \greenbullet & & & & & &
      \\\hline
      Zhang et al. \cite{zhang2024self} & & & \greenbullet & & & & & & &
      \\\hline 
      MetaAgents \cite{li2023metaagents} & \greenbullet & & & & & & & & &
      \\\hline
      Xie et al. \cite{xie2024can} & & & & \greenbullet & & & & & &
      \\\hline
      CRSEC \cite{ren2024emergence} & & & & \greenbullet & & & & \greenbullet & &
      \\\hline
      He et al. \cite{he2023homophily} & & & & & & & & & \greenbullet &
      \\\hline
      Wu et al. \cite{wu2024shall} & & & & & \greenbullet  & & & & & 
      \\\hline
      Digital Representatives \cite{jarrett2023language} & \greenbullet & & & & & & & & &
      \\\hline
      Affordable Generative Agents \cite{yu2024affordable} & \greenbullet & & \greenbullet & & & & & & \greenbullet &
      \\\hline
      Chuang et al. \cite{chuang2024wisdom} & & & & & & & & & & \greenbullet
      \\\hline 
      FUSE \cite{liu2024tiny} & & & & & & & & \greenbullet & &
      \\\hline
      HiSim \cite{mou2024unveiling} & & & \greenbullet & & & & & & & \greenbullet
      \\\hline
      OASIS \cite{yang2024oasis} & & & & & & & & \greenbullet & & \greenbullet
      \\\hline
      Sreedhar \& Chilton \cite{sreedhar2024simulating} & & & & & \greenbullet & & & & &
      \\\hline
      LMAgent \cite{liu2024lmagentlargescalemultimodalagents} & \greenbullet & & \greenbullet & & \greenbullet & & & & &
      \\\hline
      Gu et al. \cite{gu2025largelanguagemodeldriven} &  & &  \greenbullet & & & & & & & \greenbullet
      \\\hline
      Orlando et al. \cite{orlando2025generativeagentbasedmodelingreplicate} &  & &  & & & & & & & \greenbullet
      \\\hline
      Wang et al. \cite{wang2025user} & \greenbullet & &  & & & & & & & \greenbullet
      \\\hline
      Ferraro et al. \cite{Ferraro2025155} & & & \greenbullet  & & & & & & & \greenbullet
      \\\hline
      Zheng et al. \cite{Zheng202563} & & &  & & & & & & & \greenbullet
      \\\hline
      Total & 8 & 2 & 10 & 5 & 7 & 2 & & 8 & 5 & 10
      \\[.5ex]
      \multicolumn{2}{c}{} &
      \multicolumn{4}{c}{\bfseries Individual} &&&
      \multicolumn{2}{c}{\bfseries Group} &
      \\
      
    \end{tabular}%
    \caption{Classification of the reproduced social phenomena for the 35 papers}
    \label{tab:rq1}
    \end{table*}
    \kern19.5mm % manually with the help of the next \vrule
    % \vrule height 50mm
  \endgroup

\subsubsection*{Profile Alignment}
An important aspect of human behavior which modelers seek to reproduce is the way human actions are shaped by their memories, personality traits, and experiences. We may refer to this as \textit{profile alignment}. Systematically validating profile alignment however represents a challenge, as it not clear what constitutes the `correct' agent action based on their profiles and specific situations. Ground truth is often challenging or nearly impossible to access, as it would require comparing the model agent's behavior against the specific person being modelled. Park et al \cite{park2023generative} assess the validity by interviewing agents on their self-knowledge, memory, plans, reactions, and reflections, and judging the answers 'believability' based on the alignment with the personality and memories of the agents as well as with the environment. As Yu et al. \cite{yu2024affordable} build upon these Generative Agents, they also measure the believability of their Affordable Generative Agents. Wang et al. \cite{wang2025user} also search for believability of their agents' behavior and functioning of the memory system. Hua et al \cite{hua2023war}, which use LLM agents to simulate the interactions between conflicting countries,  evaluate the congruence, stability, and rationality of actions with respect to the country profile in a war situation using expert assessment. Altera.ai \cite{al2024project} deployed LLM agents in a Minecraft session, and sought to validate profile alignment by monitoring the specialization into distinct roles that should align with the profile of the agent, which in turn should align with the agent's actions. Jarrett et al \cite{jarrett2023language} create `digital representatives' of individuals, which are validated based on their alignment with the preferences of the individuals they substitute in collective decision-making scenarios. Liu et al. \cite{liu2024lmagentlargescalemultimodalagents} validate the alignment of the agents' shopping behavior with the profile, context and established rules and expectations.

\subsubsection*{Emotional alignment}
Two papers sought to use LLMs to simulate human emotions. Gao et al \cite{gao2023s} sought to predict the emotions of social media users towards a particular event based on three levels: calm, moderate, and intense. Wang et al \cite{wang2023humanoid} develop a platform for modeling human behavior, including basic needs, social relationships, and emotions. The paper validates the emotions expressed by agents for different activities through comparison with human annotation. 

\subsubsection*{Conversations and content generation}
In many papers, the agents can interact with each other through conversation. Simulations of, for instance, social media platforms often seek to validate that the content produced matches the content produced by human participants. Park et al \cite{park2022social} create 'SimReddit' -- a synthetic version of the social media platform Reddit -- and evaluate whether human participants can distinguish the synthetic conversations from human conversations. \cite{yu2024affordable} leveraged the same strategy to validate the conversations of LLM agents interacting in a virtual town. Gu et al. \cite{gu2025largelanguagemodeldriven} took a turn to quantitative text measures when comparing synthetic and real-world conversations, measuring the cosine similarity between the generated text and human text embeddings, and Gao et al. \cite{gao2023s} extends this with the Perplexity score of the generated text as well. Ferraro et al. \cite{Ferraro2025155} try to find out whether generative agents are able to produce similar keywords, interests and content compared to humans on social media. Mou et al \cite{mou2024unveiling} focuses instead on the agents' stances and the alignment of the generated content with real-world Twitter data. They furthermore categorize different types of content -- e.g., 'call for action', 'sharing of opinion', or 'reference to a third party' -- and compare whether these match the Twitter data. 

Others are more concerned with the humanness of the conversation itself, as opposed to the similarity of the content. Wang et al \cite{wang2023humanoid} assessed conversational realism based on whether engaging in a conversation brought agents closer to one another. Zhang et al \cite{zhang2024speechagents} design a system to simulate human communication, validating the performance based on consistency with the scenario and characters, and the quality and logical coherence of the script content. Zhang \textit{et al.} \cite{zhang2024self} tried to validate whether their model is able to generate human-like dialogues by scoring it on the naturalness, empathy, interestingness and humanness of the conversations, as well as the agent's ability to choose a fitting dialogue strategy. Similarly, Liu et al. \cite{liu2024lmagentlargescalemultimodalagents} compared the content of their agents with human content on naturalness and expressiveness. 

\subsubsection*{Social awareness and social intelligence}
Social awareness refers to the collective consciousness shared by individuals in a society \cite{schlitz2010worldview}. Validation may focus on the agents' ability to adhere to social norms and rules. Simulations may also focus on the related phenomenon of social intelligence, in which agents seek to reproduce the human ability to interpret actions of others. 

Altera.ai \cite{al2024project} validate the capacities of the agents to understand both themselves and others. The authors assess whether agents were able to accurately deduce the sentiment of others and react to changing social cues. Li et al. \cite{li2023theory} test whether agents can display a `Theory of Mind', that is, whether they can reason about the concealed mental states of other agents. Zhou et al \cite{zhou2023sotopia} simulate  interactions in various social scenarios to measure the social intelligence of agents based on a newly created benchmark, SOTOPIA-EVAL, which scores social intelligence based on believability, knowledge, keeping or revealing secrets, relationships, social rules, and financial and material benefits. Xie \textit{et al.} \cite{xie2024can} measured the extent to which agents exhibit human trust behavior, based on anticipating the actions and thoughts of other players in games like the Trust Game and the Dictator Game. Ren et al \cite{ren2024emergence} focus on the emergence of social norms and how they are incorporated in the agents' planning and actions in a sandbox society.

\subsubsection*{Decision-making and reasoning}
Reproducing human decision making or reasoning is a common focus of generative ABMs. Gao et al \cite{gao2023s}, for instance, compare agents' decisions to take specific actions on social media -- such as sharing, posting new content, or doing nothing -- against real user decisions. Kaiya et al \cite{kaiya2023lyfe} evaluated agents based on their autonomy and social reasoning capabilities in three experimental scenarios: a murder mystery, a high school activity fair, and a patient-in-help scenario. Williams et al. \cite{williams2023epidemic} examined whether agents were able to make the decision to stay at home or go into quarantine in a COVID-19 like epidemic scenario. Altera.ai \cite{al2024project} assessed agents' ability to reason about societal rules. Wu et al. \cite{wu2024shall} explore whether agents were able to make adaptive decisions without explicit directions, focusing on three scenarios where agents can only reach an optimal outcome when working together in a competitive context. Sreedhar and Chilton \cite{sreedhar2024simulating} simulate the Ultimatum Game with LLM agents to simulate and replicate human strategic reasoning. Liu et al. \cite{liu2024lmagentlargescalemultimodalagents} simulate user shopping behavior and evaluate on the shopping decisions made by the agents.

\subsubsection*{Opinions and attitudes}
Two papers focused on using generative agents to simulate human attitude or opinions. Validations of this category can be focused on replicating the change of attitude or opinion in reaction to a certain event. Gao et al \cite{gao2023s} validate their agents' attitudes (positive/negative) towards posts on social media against real-world social media data. Wang et al \cite{wang2023humanoid} tested whether they are able to predict if certain activities satisfy basic needs of fullness, social, fun, health, and energy, thus measuring the attitude towards events and the effect on the individual agents.

\subsubsection*{Network propagation}
Network propagation was a key focus of study, with 8 out of 29 papers carrying out some form of validation of network propagation. The validation of these systems are often concerned with measuring the spread of something, like information or attitudes, and comparing this with real-world human behavioral patterns.

Park et al \cite{park2023generative} focused on the diffusion of information through the network. Gao et al \cite{gao2023s} measure three forms of propagation on social media: information, emotion, and attitudes. Li et al. \cite{li2024large} and Liu et al \cite{liu2024skepticism,liu2024tiny} simulate the spread of misinformation in a social media network structure. \cite{yang2024oasis} reported on the spread of all forms of information. Altera.ai \cite{al2024project} track the spread of cultural memes and religion in a network, and Ren et al \cite{ren2024emergence} seek to model the spread of social norms through a social network. 

\subsubsection*{Network structure}
The structure of the network is mainly determined by the formation of relationships between agents. Here, the goal of the validation is to make sure that the resulting network characteristics match real-world network properties. De Marzo et al \cite{de2023emergence}, for instance,  evaluate whether the resulting network structure from their simulation was scale-free, which is a near-universal property of networks emerging on social media.  He et al \cite{he2023homophily} examine whether the networks produced on their synthetic social media network, Chirper.ai, show the property of homophily, i.e., the tendency of agents to be more connected with others that are similar to themselves. Park et al \cite{park2023generative} and Yu et al \cite{yu2024affordable} both examine the evolution of social networks in their sandbox town simulation, using metrics such as the degree of nodes, and the density of the network. Hua et al. \cite{hua2023war} assess the historical correctness of the formed alliances and war declarations between countries in a war situation. Gu et al. \cite{gu2025largelanguagemodeldriven} try to measure the formation of echo chambers with measures like the modularity, network density, average path length and the clustering coefficient of the network.

\subsubsection*{Social dynamics}
This final category focuses on papers that seek to reproduce social dynamics, like a flock of starlings collectively forming a fluid cloud. This category encompasses group behavior that emerges from individual actions. 

Zhang et al \cite{zhang2023exploring} claim to find conformity, consensus reaching, and group dynamics behavior in their agent society. The agents in Chen et al \cite{chen2023agentverse} exhibit volunteering, conformity, and destructive behavior when trying to survive in Minecraft. Wang et al. \cite{wang2025user} also find conformity behavior, next to signs of an 'information cocoon'. A similar phenomenon, the formation of echo chambers, was also validated in various papers \cite{gu2025largelanguagemodeldriven, Ferraro2025155, Zheng202563}. Chuang et al. \cite{chuang2024wisdom} simulate the phenomenon of the Wisdom of Partisan Crowds, where the group average moves closer to the ground truth when exposed to the average belief of the partisan group, despite differences in initial belief. Mou et al \cite{mou2024unveiling} measure the attitude distribution and average attitude of users on a simulated Twitter platform, HiSim, using the deviation from the mean (bias), the standard deviation from the mean (diversity) and the similarity measured with Dynamic Time Warping and Pearson correlation. Yang et al \cite{yang2024oasis} replicate the `herd effect' on Reddit, by testing whether an initial like or dislike of a comment results in other agents following the behavior. They managed to reproduce the effect, but found that the effect was stronger among their generative agents than among human, likely as humans possess a stronger critical mind that reduces the herd effect. Orlando et al. \cite{orlando2025generativeagentbasedmodelingreplicate} measure the replication of the 'friendship paradox', the phenomenon that individuals on average have fewer friends than their friends.

\subsection{Validation Techniques}
We turn now to the question of the validation approach taken in the papers. Table \ref{tab:rq2} provides an overview of the validation techniques used. For the classification of the papers, we distinguish between primary and secondary techniques.  The primary techniques are those that are  most important in the validation process, marked in green in the table. The secondary techniques are supporting evidence of lesser importance, marked in orange in the table. 

We will here discuss the techniques used in order, drawing on examples of their use.

\begingroup
    \renewcommand*{\arraystretch}{1.5}%
    \definecolor{tabred}{RGB}{230,36,0}%
    \definecolor{tabgreen}{RGB}{0,116,21}%
    \definecolor{taborange}{RGB}{255,124,0}%
    \definecolor{tabbrown}{RGB}{171,70,0}%
    \definecolor{tabyellow}{RGB}{255,253,169}%
    \newcommand*{\redtriangle}{\textcolor{tabred}{\ding{115}}}%
    \newcommand*{\greenbullet}{\textcolor{tabgreen}{\ding{108}}}%
    \newcommand*{\orangebullet}{\textcolor{taborange}{\ding{108}}}%
    \newcommand*{\orangecirc}{\textcolor{taborange}{\ding{109}}}%
    \newcommand*{\headformat}[1]{{\footnotesize\bfseries#1}}%
    \newcommand*{\vcorr}{%
      \vadjust{\vspace{-\dp\csname @arstrutbox\endcsname}}%
      \global\let\vcorr\relax
    }%
    \newcommand*{\HeadAux}[1]{%
      \multicolumn{1}{@{}r@{}}{%
        \vcorr
        \sbox0{\headformat{\strut #1}}%
        \sbox2{\headformat{Complex Data Movement}}%
        \sbox4{\kern\tabcolsep\redtriangle\kern\tabcolsep}%
        \sbox6{\rotatebox{45}{\rule{0pt}{\dimexpr\ht0+\dp0\relax}}}%
        \sbox0{\raisebox{.5\dimexpr\dp0-\ht0\relax}[0pt][0pt]{\unhcopy0}}%
        \kern.75\wd4 %
        \rlap{%
          \raisebox{.25\wd4}{\rotatebox{45}{\unhcopy0}}%
        }%
        \kern.25\wd4 %
        \ifx\HeadLine Y%
          \showthe\dimen0 % Debugging: Check the computed width
          \dimen0=\dimexpr\wd2+.5\wd4\relax
          \rlap{\rotatebox{45}{\hbox{\vrule \vrule width 3cm height .4pt}}}%
        \fi
      }%
    }
    \newcommand*{\head}[1]{%
  \HeadAux{\def\HeadLine{Y}#1}%
  }
    \newcommand*{\headNoLine}[1]{\HeadAux{\global\let\HeadLine=N#1}}%
    \noindent
    \begin{table*}[]
    \footnotesize
    \begin{tabular}{%
      >{\bfseries}lc|
      *{2}{c|}c>{\quad}c|
      *{2}{c|}c>{\quad}c
    }%
      &
      \head{Human(-like) Judgment} &
      \head{Social Patterns} &
      \head{Other Models} &
       &
      \head{Human Generated} &
      \head{Internal Consistency} &
      \head{Other Models} &
      &

      \\
      \sbox0{S}%
      \rule{0pt}{\dimexpr\ht0 + 2ex\relax}%
      Generative Agents \cite{park2023generative} & \greenbullet & \orangebullet
       &  & & & &  &
      \\\hline
      WarAgent \cite{hua2023war} & \greenbullet &
      & & & \greenbullet & \orangebullet & &
      \\\hline
      Social Simulacra \cite{park2022social} & \greenbullet &
      &  & & \orangebullet &&&
      \\\hline
      S3 \cite{gao2023s} & & & && \greenbullet &  & 
      \\\hline
      De Marzo et al. \cite{de2023emergence} & & \greenbullet & & & & \orangebullet & &  \\\hline
      Humanoid Agents \cite{wang2023humanoid} & & & && \greenbullet &  & \\\hline
      Lyfe Agents \cite{kaiya2023lyfe} & & \greenbullet & & & & & & \\\hline
      Zhang et al. \cite{zhang2023exploring} & & \greenbullet & & & & & & \\\hline
      AgentVerse \cite{chen2023agentverse} & & \greenbullet & & & & & & \\\hline
      Williams et al. \cite{williams2023epidemic} & & \greenbullet & & & & & & \\\hline
      Project Sid \cite{al2024project} & & \greenbullet & & & & & & \\\hline
      Li et al. \cite{li2023theory} & \greenbullet & \orangebullet & & & & & & & \\\hline
      Li et al. \cite{li2024large} & & \greenbullet & & & & & & \\\hline
      SpeechAgents \cite{zhang2024speechagents} & \greenbullet & & & & & & \orangebullet & \\\hline
      FPS \cite{liu2024skepticism} & & \greenbullet & & & & & &\\\hline
      SOTOPIA \cite{zhou2023sotopia} & \greenbullet & & && \orangebullet & & & \\\hline
      Zhang et al. \cite{zhang2024self} & \greenbullet & & && \greenbullet & & & \\\hline
      MetaAgents \cite{li2023metaagents} & \greenbullet & & & & & & & \\\hline
      Xie et al. \cite{xie2024can} & \greenbullet & & & && & & \\\hline
      CRSEC \cite{ren2024emergence} & \greenbullet & \orangebullet & &&  & & & & \\\hline
      He et al. \cite{he2023homophily} & & \orangebullet & & & & \greenbullet & &  \\\hline
      Wu et al. \cite{wu2024shall} & & \greenbullet & & & \orangebullet & & & \\\hline
      Digital Representatives \cite{jarrett2023language} & \orangebullet & & & & \greenbullet & & & \\\hline
      Affordable Generative Agents \cite{yu2024affordable} & \greenbullet & & \greenbullet & & & & \orangebullet & & \\\hline
      Chuang et al. \cite{chuang2024wisdom} & & & & & \greenbullet & & & \\\hline
      FUSE \cite{liu2024tiny} & & \greenbullet & & & & & & \\\hline
      HiSim \cite{mou2024unveiling} & & \orangebullet & & & \greenbullet & & \greenbullet & \\\hline
      OASIS \cite{yang2024oasis} & & \orangebullet & & & \greenbullet & & & \\\hline
      Sreedhar \& Chilton \cite{sreedhar2024simulating}  & & & & & \greenbullet & \orangebullet & & 
      \\\hline
      LMAgent \cite{liu2024lmagentlargescalemultimodalagents}  & \greenbullet & & \orangebullet & & \greenbullet & & & 
      \\\hline
      Gu et al. \cite{gu2025largelanguagemodeldriven}  & & & \orangebullet & & \greenbullet & & & 
      \\\hline
      Orlando et al. \cite{orlando2025generativeagentbasedmodelingreplicate}  & & \greenbullet &  & &  & & & 
      \\\hline
      Wang et al. \cite{wang2025user}  & \orangebullet & \greenbullet & \orangebullet & & \orangebullet & & & 
      \\\hline
      Ferraro et al. \cite{orlando2025generativeagentbasedmodelingreplicate}  & & \greenbullet &  & & \greenbullet & & & 
      \\\hline
      Zheng et al. \cite{Zheng202563}  & & \greenbullet &  & &  & & & 
      \\\hline
      Total (main technique) & 12 & 14  & 1 && 12 & 1 & 1
      \\\hline
      Total (secondary technique) & 2 & 6 & 3 & & 4 & 3 & 2
      \\[.5ex]
      \multicolumn{1}{c}{} &
      \multicolumn{3}{c}{\bfseries Subjective} &&
      \multicolumn{4}{c}{\bfseries Objective} &

    \end{tabular}%
    \caption{Classification of the validation techniques for the 35 papers.}
    \label{tab:rq2}
    \end{table*}
    \kern19.5mm % manually with the help of the next \vrule
    % \vrule height 50mm
  \endgroup

\subsubsection*{Validation based on human or human-like judgment}
One of the most widely used techniques for validation of generative agents is to draw on external or internal judges. The judges can be 1) experts (often in the form of the authors themselves), 2) crowd-workers, or 3) LLMs. For the first category, Hua et al \cite{hua2023war} use field experts to evaluate the congruence and consistency of actions taken by countries with respect to their profile. Li et al \cite{li2023metaagents} use the authors' own judgment to assess the performance of their MetaAgents to identify suited job-seekers, create a workflow and align with the given role. Wang et al. \cite{wang2025user} make use of humans to asses the believability of the chatting behavior and memory module of the agents.

While the use of crowd-workers has been widely criticized both on ethical grounds, for low performance, and for their now wide-spread use of LLMs, crowd-workers remain widely used for validation, often hired through Amazon MTurk. Park et al \cite{park2023generative} recruited crowd-workers to score the agents on `believability'. Crowd-workers were also used in Park et al \cite{park2022social}, to distinguish between synthetic conversations and real conversations for repopulated subreddits. Human annotators were used to evaluate whether generative agents answer Theory of Mind questions correctly in Li et al \cite{li2023theory}.  Several other papers also employ crowd-workers to evalaute whether humans can distinguish human and generative agents on various tasks \cite{ren2024emergence,zhou2023sotopia,zhang2024self}.

While using LLMs to validate the realism of LLMs appears self-evidently problematic, the practice has become increasingly widespread as a means of validation. Zhou et al \cite{zhou2023sotopia} use GPT-4 to validate their model realism. Zhang et al \cite{zhang2024speechagents} use ChatGPT to assess the consistency and coherence of the conversational content. Jarrett et al \cite{jarrett2023language} use an LLM to compare  generated critique to a ground-truth. Liu et al. \cite{liu2024lmagentlargescalemultimodalagents} use both humans and GPT-4 to evaluate the behavior of agents on among other believability, social influence and the naturalness and expressiveness of the content. Yu et al \cite{yu2024affordable} use GPT-4 to discern whether responses originated from AI or a human, despite previous research suggesting that LLMs may be inappropriate for such tasks \cite{bhattacharjee2024fighting}.

\subsubsection*{Validation against well-known social patterns}
Validation can also be carried out by matching the dynamics of the model with well-established real-world patterns. This means that no specific empirical data need to be collected for comparison, as the comparison is made against known social patterns. 

De Marzo et al. \cite{de2023emergence} examine whether the networks generated show the property of having scale-free degree distributions, which is a well-known and easily measurable property of social media networks. Li et al \cite{li2024large} show that the effect of different personalities and agent attributes on  agents' propensity to share fake news matches earlier studies. Liu et al \cite{liu2024skepticism} conclude that their finding that fake political news propagates faster than other fake news topics is consistent with previous research. In addition, they compare the relationship between the Big Five personalities on agents' propensity to spread misinformation with prior empirical research on the spread of misinformation. Orlando et al. \cite{orlando2025generativeagentbasedmodelingreplicate} use the degrees of agents to validate the known phenomenon of the 'friendship paradox'. 

While some social patterns, such as degree distributions, are easily quantified, many known social patterns are qualitative rather than quantitative, meaning that authors often need to argue for the plausibility of the model in a more narrative form -- with varying levels of persuasiveness and rigor. For instance, Zhang et al. \cite{zhang2023exploring} observe dynamics in their model that they argue corresponds to well-studied behavior in social psychology, such as conformity, consensus-reaching, and group dynamics. Chen et al \cite{chen2023agentverse} similarly observe social dynamics that they argue match established theories within social psychology. Williams et al \cite{williams2023epidemic} argue that the actions taken by agents in response to disease outbreak -- such as staying at home and quarantining -- share some similarities to the  behavior observed during the COVID-19 pandemic. Yang et al \cite{yang2024oasis} observe the emergence of group polarization in their model, which they argue matches behavior observed in humans. 

Kaiya et al \cite{kaiya2023lyfe} and Altera.ai \cite{al2024project} take an even more loose approach, merely observing the simulated system and claiming that it matches human behavior. Ren et al \cite{ren2024emergence} take a case study approach, observing their system and interpreting the dynamics as representing the adoption of norms and formation of social conflicts, claiming that it thereby matches real-world social dynamics. Wu et al \cite{wu2024shall} find that agents agree to cooperate without explicit directions and report on performance metrics, which were compared against a study on humans in only one of the three experiments. Yu et al \cite{yu2024affordable} and Park et al \cite{park2023generative} argue that the formation of relationships and the spread of information through the network matches expected social dynamics. Mou et al \cite{mou2024unveiling}, Zheng et al \cite{Zheng202563} and Ferraro et al \cite{Ferraro2025155} remark that their systems reproduce the known social pattern of echo chambers. Wang et al. \cite{wang2025user} similarly find signs of an 'information cocoon' and conformity behavior.

\subsubsection*{Validation against similar models}
Generative ABMs can also be validated through comparison with previous models. The catch here is that it is implied that the previous models are already rigorously validated, which circles back to the validation problem that this paper addresses. As a consequence, authors can conclude on the performance compared to the other model, but the alignment with human behavior depends heavily on the quality of the other model. These comparisons again vary in whether it is carried out in a rigorous quantitative way or in a more subjective way that is argued for in narrative form. 

\cite{zhang2024speechagents} compare the generated conversations of their SpeechAgents with a previous single-agent model. Mou et al \cite{mou2024unveiling} and Gu et al \cite{gu2025largelanguagemodeldriven} systematically compare their social media simulations with conventional non-generative ABMs. Liu et al \cite{liu2024lmagentlargescalemultimodalagents} compare the purchasing behavior of their agents to different filtering algorithms and two other agent based approaches. This system is in turn used by Wang et al \cite{wang2025user} in a comparison with their approach to recommender agents. 

Yu et al \cite{yu2024affordable} offers an example of a more subjective comparison, comparing their model with Park et al's \cite{park2023generative} Generative Agents. The main mode of validation in both papers was the `believability' of the agents' responses to a set of interview questions. Yu et al \cite{yu2024affordable}, however, offers no quantitative comparison between the two models, but merely includes the responses of agents, concluding that their version did not obviously impair the believability of the agents compared to the original. 

\subsubsection*{Validation against human-generated data}
One common technique for validation is to quantitatively compare the synthetic data produced by the model with `real-world' data, for instance collected from digital media or from human annotation. This technique is particularly common within social media simulations. Gao et al \cite{gao2023s}, Mou et al \cite{mou2024unveiling}, Yang et al \cite{yang2024oasis}, Gu et al \cite{gu2025largelanguagemodeldriven} and Ferraro et al \cite{Ferraro2025155} all based their validation on social media datasets on comparison with data collected from platforms such as Twitter or Reddit. Used metrics include the speed and scale of the spread of information through the network, the stance and attitude of users towards certain events and the similarity of the generated content. In a non-social media system, Hua et al \cite{hua2023war} use historical data to qualitatively assess the accuracy of their war simulation, such as war declarations or alliance formation. Liu et al \cite{liu2024lmagentlargescalemultimodalagents} use a dataset of Amazon products and reviews, use a part of the purchase history for the initialization of agents and keep the rest as a ground truth for comparison to see to what degree agents make the same purchases.

Another common approach is to use simulations to reproduce experiments that have previously been carried out with human participants. Papers have been found to compare the results of generative agents on trust games \cite{xie2024can}, the Ultimatum Game \cite{sreedhar2024simulating} and tests to measure the effect of the Wisdom of Partisan Crowds \cite{chuang2024wisdom} to previous studies with human subjects.

Finally, Wang et al \cite{wang2023humanoid} compare classifications of human annotators with classifications made by the generative ABM on emotions. Zhang et al \cite{zhang2024self} use human annotations as a ground truth to validate generative agent choices of dialogue strategies. Jarrett et al \cite{jarrett2023language} validate the behavior of agents as digital representatives against an annotated dataset of human opinions and critiques in combination with demographic information. 

\subsubsection*{Validation based on internal consistency}
Some researchers seek to validate their models based on their internal consistency, coherence, and stability, for example, through sensitivity analysis. The goal may be to show that the agents' behavior is stable to changes in system parameters and perturbations. For example, Hua et al \cite{hua2023war} experiment with injecting counterfactual information and de-anonymization of countries to compare with the base approach to see if the results were as expected. De Marzo et al \cite{de2023emergence} use different prompts for their language model to see if the network structure instead converged to an expected structure, such as a random graph. He et al \cite{he2023homophily} use statistical tests on network metrics to show that the network data can be divided into distinct communities. Sreedhar and Chilton \cite{sreedhar2024simulating} use simulations of agents with different personality traits to evaluate whether the differences between the models align with expectations. While these form of internal tests are relatively common, the extent to which they can claim to represent a plausible validation of the model, rather than just a verification of its basic functioning, is however debatable. 

\section{Discussion} 
\label{sec:discussion}
Based on our review of the way generative ABMs have been employed and validated in the literature, we now turn now to our central question of how and whether the field is sufficiently addressing the long-standing challenges of validation in ABM, as well as the new challenges that emerge from the use of LLMs.

As we have seen, the most common validation technique is to simply employ on-the-face `believability': 15 out of 35 papers are validated solely through subjective validation, and 22 out of 35 use subjective validation as their only primary technique. 14 papers use the judgment of human `experts', crowd-workers, or LLMs as evaluators. At its best, this consists of the researchers themselves looking at the simulated system and arguing -- with varying rigour -- that it shares some similarity with human behavior or a social system. This form of validation can also involve using MTurk workers, despite the fact that the low quality of such validation data at this point has been well established \cite{karpinska2021perils}. At its worst, this form of validation consists of simply asking an LLM to evaluate its own believability. Various papers have doubted the qualities of LLMs for evaluation purposes \cite{wang2023large, panickssery2024llm, polakova2024examining}. On the other hand, some have argued that LLMs as evaluators are as good as humans \cite{chan2023chateval, zhou2023sotopia}, this, however, should be taken as an argument against using human judgment in validation, rather than as an argument for using LLMs. Humans -- in particular those without training -- are famously poor at judging whether a material is produced by AI, and it is arguably relatively simple to set up an experiment in such a way as to generate a positive outcome with even low-quality results -- for instance by asking whether a given text is generated by AI or by humans, rather than by allowing side-by-side comparison between the synthetic and real-world data.

When studies do carry out more rigorous quantitative comparison between human-generated and synthetic data, it often quickly becomes clear that the two are substantially less similar than the on-the-face believability may suggest. The style of writing, which tends to be the focus of such comparisons, between zero-shot LLMs and human conversations tends to be quite different: LLM responses are longer, more polite, articulate, and respectful \cite{zhang2024self,mou2024unveiling}. While humans struggle to tell the difference, there are generally substantial syntactical differences between human-generated and synthetic text.

Yet, from the perspective of the literature on ABM validation, even the few studies that carry out more rigorous quantitative validation against real-world data show some fundamental issues. As mentioned, the notion of operational validity highlights that the purpose of validation is for the model to match the underlying mechanisms of the object over the domain of the model's intended applicability \cite{sargent2010verification}. In the case of generative ABMs, it is rarely clear that the model aspects being validated are at all relevant for the model's mechanisms. Take, for instance, the models seeking to simulate social media platforms. The purpose of such simulation tends to be to test the effect of affordances or algorithms on social dynamics on social media. Yet, the validation tends to focus on whether the synthetic text produced shares some superficial syntactical features with real-world social media text. Arguably, unless the algorithms being tested happen to leverage syntactical features of the text, such aspects are irrelevant for operational validity, which would rather depend on how and when users react to one another, and how these actions in turn interact with the affordances and algorithms of the platform. The sufficiency of the aspects that are being validated for the dynamics of the model must be clearly evidenced. 

Operational validity would moreover often require validating profile alignment for the individuals being simulated. In other words, it would be necessary not only to test that the agent is acting consistently with itself through the simulation, but that it is acting in the same way as the person being simulated -- generally, in the included studies, a person described through a prompt persona. How to carry out such validation is however an open question, as rigorous persona alignment would require calibrating and validating against a specific individual and their actions within situation being captured by the model. Doing so would in most cases imply either collecting specific data on individuals through linked surveys or human-participant experiments, or seeking to create what we may call `digital human twins' by calibrating on digital trace data -- in turn implying challenging ethical questions. 

In existing studies, however, nearly all studies use zero-shot models without finetuning. `Calibration' hence consists of using prompt engineering to achieve on-the-face believability. As a result, it is simply assumed that LLMs will accurately model individuals -- despite that it has been shown that the models are poor at representing social groups, not least due to often problematic social bias \cite{boelaert2024machine}. While some papers do mention the problems of social bias in LLMs, for instance finding that the agents display gender biases in the simulation \cite{xie2024can}, the more fundamental question of how this impacts calibration of the model against human behavior is not raised. 

The central feature of ABMs -- their inherent flexibility that simultaneously affords their great versatility but also makes them challenging to calibrate, validate, reproduce, and compare, and that make it nearly impossible to achieve cumulativity of findings -- does not appear to have in any way been alleviated by the introduction of LLMs, but -- if anything -- exacerbated. While some of the papers do recognize the challenge implied by the lack of a shared validation framework or benchmarks \cite{de2023emergence, kaiya2023lyfe, chen2023agentverse, wu2024shall,yu2024affordable}, and some even propose new frameworks \cite{al2024project,li2023metaagents,chuang2024wisdom, liu2024skepticism}, such suggestions have been long-standing features of the ABM literature, and the challenge of negotiating the tradeoffs between versatility and standardization is not made easier by the introduction of LLMs. Historically, such frameworks have either seen limited re-use, or do not constraint ABMs enough to achieve standardization. 

The introduction of LLMs also supercharges the long-standing challenge of the high computational costs associated to ABMs. While large simulations were always costly, the costs of conventional models appear modest compared to those of LLM-based social simulation. Some studies attempt to mitigate these costs by optimizing agent implementations \cite{kaiya2023lyfe, yu2024affordable}, leveraging distributed processing \cite{yang2024oasis} or replacing less critical agents with traditional ABMs \cite{mou2024unveiling}. However, these strategies do not change the fundamental reality: generative ABMs are orders of magnitude more computationally demanding than conventional ABMs—already criticized for their high costs. As a result, none of the existing generative ABM studies conduct the necessary sensitivity analysis to assess model reliability and identify key parameters. Even more concerning, most rely on a \textit{single model run}. Given the inherent stochasticity of generative ABMs, this approach is highly problematic -- akin to drawing conclusions from a single-case study.

% Bias is one of the most frequently discussed challenges in the included papers. First, there is the already mentioned social bias, which gets noticed but not tackled \cite{xie2024can, he2023homophily, zhou2023sotopia, jarrett2023language, ren2024emergence}, but there is also the selection bias. Agents where found to resemble facts about historical figures or names, such as "Adam Smith" or "COVID" \cite{park2023generative, williams2023epidemic, de2023emergence, hua2023war}. Going a step further, there were concerns that LLMs could be trained on social media conversations \cite{park2022social}, historical events \cite{hua2023war} or strategies for games like the Ultimatum Game \cite{sreedhar2024simulating}. This would mean that agents simply replicated training data instead of adapting to the situation.

% Finally, LLMs suffer from hallucinations -- the generation of nonsensical information. Examples are common in the literature. For example, agents were observed to answer with information that was not in their memory \cite{park2023generative, li2023theory}, showed coherence problems; they claimed to perform actions but were not actually doing it \cite{al2024project}, or performed actions that explicitly not allowed by the modelers \cite{mou2024unveiling}. Hallucinations are commonly battled by the introduction of internal sanity check mechanisms, often resulting in the problematic situation where LLMs fact-check the output of other LLMs.

\section{Conclusion}
\label{sec:conclusion}
Generative ABM is an exciting and quickly growing field, making use of the capacity of LLMs to mimic human behavior. This paper has situated generative ABMs in the longer history of, and debate surrounding, agent-based modeling. The paper has reviewed the field to assess whether and how the now-rejuvenated field deals with the long-standing challenges that have historically limited the use of ABMs in the social sciences. 

Our review has shown that the rapidly growing field is drawing on LLMs to model a wide range of human behavior, ranging from network dynamics to social media simulations. LLMs enable models to appear much more `realistic', and open up for exploring aspects of human behavior that have previously been nearly impossible to model. 

However, based on the review of the recent literature, we have also found that the addition of LLMs to ABMs does little to address the challenges with which the field has struggled, often exacerbating rather than resolving the long-standing issues limiting the use of ABMs. 

The lack of rigor in relation to issues such as validation is partly excused by the fact that the field of generative ABMs is still in a stage of early experimentation, exploring the potential of using LLMs to reproduce aspects of human behavior through proof-of-concept models. This begs the question of whether field will be able to transition from such toy models to the type of rigorous modeling needed to contribute productively to social scientific theory. 

To do so, it would be necessary to evidence operational validity \cite{sargent2010verification}: the model must be convincingly shown to capture some existing social mechanism. For conventional ABMs, there have been two chief ways of achieving operational validity.

The first way is to employ carefully calibration and validation to achieve model realism \cite{edmonds2004kiss}. LLMs appear to naturally afford this route, as they enable seemingly more realistic representations of human reasoning and decision-making. However, the introduction of LLMs do not appear to resolve the issues that made such validation and calibration challenging for ABMs, such as their inherent flexibility and lack of standardization and comparability. LLMs furthermore add to this the challenging task of developing calibration and validation procedures to culturally align the agents with real-world individuals, for instance showing that the model is not representing minorities by reproducing problematic stereotypes. 

The second way to achieve operational validity is to focus on capturing highly simple -- and therefore robust and generalizable -- emergent phenomena. Many of the most insightful ABMs are highly simplistic thought-experiments that throw light on a minimal representation of a phenomenon. As such, these models do not necessarily ``aim to provide an accurate representation of a particular empirical application. Instead, the goal of agent-based modeling is to enrich our understanding of fundamental processes that may appear in a variety of applications'' \cite[25]{axelrod1997advancing}. Scholars who take this approach to ABMs are ``much more concerned with theoretical development and explanation than with prediction'' \cite[21]{gilbert1997simulation}; ABMs offer a form of computational thought-experiments that enable exploring whether particular micro-mechanisms can produce an observed pattern. Such models are believable not because of their realism but because the emergent phenomenon that they capture is so simple and general as to be applicable to a wide range of systems. The famous Schelling segregation model \cite{schelling1971dynamic}, for example, is insightful not because the model accurately represents cities, but because the dynamics that it describes can offer insights into phenomena ranging from residential segregation to why oil separates from water. 

However, it is not clear how generative ABMs would be useful for either of these two approaches. For the former, generative ABMs are, as we have seen, exceedingly challenging to calibrate and validate against real-world data. It has yet to be shown that the models can be made to reproduce human actions in such a way as to make them social scientifically productive. For the latter, generative ABMs rely on highly complicated black-box models whose behaviors are poorly understood even on their own, and it is hence not clear how these models may be used to reveal simple emergent mechanisms. While LLMs are exceptionally simple to use, as one can instruct the agents through natural language rather than employing the often complex mathematical and highly parameterized work of conventional ABMs, they can yet be highly challenging to interpret. As Axelrod \cite[27]{axelrod1997advancing} puts it, ``If the goal is to deepen our understanding of some fundamental process, then simplicity of the assumptions is important and realistic representation of all the details of a particular setting is not''. The introduction of LLMs into these models inevitably adds complexity that may undermine their usefulness as tools for theoretical research, as it makes it more challenging to figure out why and how the model produces a given result \cite{macy2002factors}. Thought-experiments are productive because they are so simple as to constitute explanations, and it is not clear whether generative ABMs will afford such simplicity.

The issues raised by this paper beg the more fundamental question of whether there is an added value to introducing LLMs in ABMs, beyond the undeniable excitement of novelty. While generative ABMs may exacerbate rather than resolve the long-standing challenges of ABMs, there are still reasons to be optimistic about their possibilities. A range of scholars have long criticized ABMs on epistemic grounds, arguing that complexity in the social world is fundamentally different from type of complexity that ABMs enable us to study \cite{byrne2022complexity,bhaskar2014possibility, bateson2000steps,tornberg2025seeing}. Human society is not akin to a murmuration of starlings: it is characterized not by emergent patterns that stem from the mass-interaction of simple agents, but from agents that \textit{recognize} and \textit{interact} with the social context within which they are operating. In the social world, emergent patterns are not just self-organizing structures; they become named, institutionalized, and capable of exerting downward causation, shaping individual and collective behavior. A potential answer to the question of the novelty brought by generative ABMs is hence that they open the door to studying yet unexplored aspects of social emergence, such as the central role of narrative \cite{polkinghorne1988narrative} and social construction \cite{berger1966social} in the human world. However, doing so requires addressing the challenges raised in this paper, to avoid the door opening merely to a new replication crisis. 

\bibliography{sn-bibliography}% common bib file


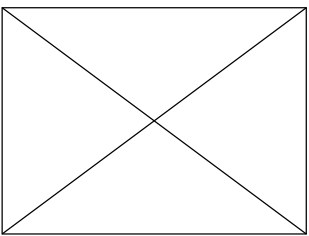
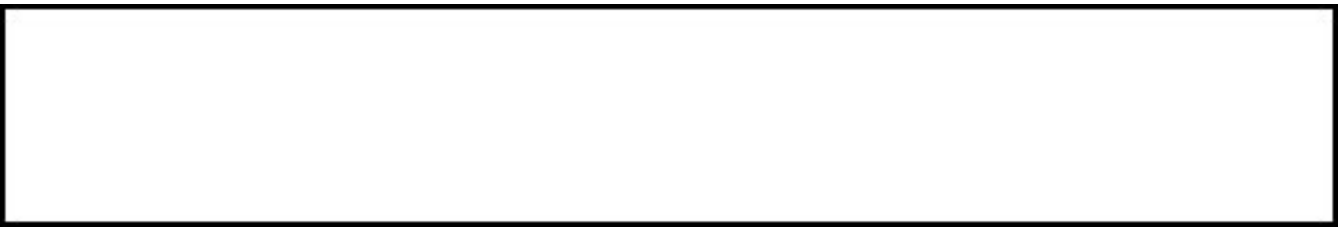
\end{document}